\documentclass[aps,showpacs,preprintnumbers,amsmath, amssymb]{revtex4}

\usepackage{float}
\usepackage{graphics,epsfig}
\usepackage{graphicx}
\usepackage{dcolumn}
\usepackage{bm}

\begin{document}
\baselineskip=0.8 cm
\title{{\bf Holographic superconductor models with $RF^{2}$ corrections}}

\author{Zixu Zhao$^{1,2}$, Qiyuan Pan$^{1,2}$\footnote{panqiyuan@126.com}, Songbai Chen$^{1,2}$\footnote{csb3752@hunnu.edu.cn} and Jiliang Jing$^{1,2}$\footnote{jljing@hunnu.edu.cn}}
\affiliation{$^{1}$Institute of Physics and Department of Physics,
Hunan Normal University, Changsha, Hunan 410081, China}
\affiliation{$^{2}$ Key Laboratory of Low Dimensional Quantum
Structures and Quantum Control of Ministry of Education, Hunan
Normal University, Changsha, Hunan 410081, China}

\vspace*{0.2cm}
\begin{abstract}
\baselineskip=0.6 cm
\begin{center}
{\bf Abstract}
\end{center}

We investigate the effect of the $RF^2$ correction on the holographic superconductor model in the background of AdS black hole. We find that, similar to the effect caused by the Weyl correction, the higher $RF^2$ correction term can make it easier for the scalar operator to condense and result in the larger deviations from the expected relation in the gap frequency. However, if a non-trivial hair for the black hole has been triggered, we observe that the $RF^2$ correction and the Weyl correction do play different roles in the behavior of the condensation, which can be used to support the existing findings.

\end{abstract}


\maketitle
\newpage
\vspace*{0.2cm}

\section{Introduction}

The anti-de Sitter/conformal field theory (AdS/CFT) correspondence, which introduced by Maldacena in 1998 \cite{Maldacena}, has provided a method to study the strong coupled field theories in a weakly coupled dual gravitational description \cite{Gubser1998,Witten}. In recent years, One of the significant achievements of this duality is that the gravitational physics has been connected with the condensed matter physics \cite{GubserPRD78}, which presents a tractable model governing superconductivity for non-standard dynamical mechanisms \cite{HartnollJHEP12}. It is found that the properties of a ($2+1$)-dimensional superconductor can indeed be reproduced in the ($3+1$)-dimensional holographic dual model in the background of AdS black hole, which requires a system that admits black holes with scalar hair at low temperature but no hair at high temperature \cite{HartnollPRL101}. Along this line, there have been accumulated interest to investigate various gravity models with the property of the so-called holographic superconductor, for reviews, see Ref. \cite{HartnollRev,HerzogRev,HorowitzRev} and references therein.

In most cases, the studies on the holographic superconductors focus on the Einstein-Maxwell theory. In order to understand the influences of the $1/N$ or $1/\lambda$ ($\lambda$ is the 't Hooft coupling) corrections on the holographic dual models, there have been a lot of works studying the holographic superconductors with the higher derivative correction related to the gravity \cite{Gregory,Pan-Wang,Ge-Wang} and the gauge field \cite{JS2010,JLQS2012,PJWPRD,JingJHEP,SDSL2012,LPW2012,BGRL2012,JPCPLB,Roychowdhury,ZPCJ2012}. Recently, an s-wave holographic dual model with Weyl corrections has been constructed in order to explore the effects beyond the large $N$ limit on the holographic superconductor \cite{Wu-Cao}. It is shown that, in strong contrast to the higher curvature corrections \cite{Gregory,Pan-Wang,Ge-Wang}, the higher Weyl corrections make it easier for the condensation to form and change the expected relation in the gap frequency. Generalizing the investigation on the holographic superconductor model in the St$\ddot{u}$ckelberg mechanism, the authors of Ref. \cite{MaCW} found that different values of Weyl correction term and model parameters can determine the order of phase transitions and critical exponents of second-order phase transitions. More recently, there seems to have developed increasing interest in investigation of Weyl corrections on the holographic dual models \cite{DMMPLA,RoychowdhuryPRD,MomeniWeyl,ZPJ2012}.

As a matter of fact, the form of the higher derivative corrections is not unique. In Ref. \cite{RCMyers}, Myers, Sachdev and Singh introduced another form of higher order corrections related to the gauge field
\begin{eqnarray}\label{RFAction}
\mathcal{L}_{RF^{2}}={\alpha}L^{2}(
R_{\mu\nu\rho\lambda}F^{\mu\nu}F^{\rho\lambda}-4R_{\mu\nu}F^{\mu\rho}F^{\nu}
_{~\rho}+RF^{\mu\nu}F_{\mu\nu}),
\end{eqnarray}
where $\alpha$ is the $RF^{2}$ coupling parameter. They argued that this correction term arises from the Kaluza-Klein reduction of five dimensional Gauss-Bonnet gravity and can produce second-order equations of motion for both the gauge field and metric in any general background \cite{RCMyers}. Cai and Pang studied the holographic properties of charged black holes with $RF^{2}$ corrections and observed that this correction will affect the DC conductivity \cite{CaiPang}. Considering this new form of higher derivative corrections, we wonder how the properties of the holographic superconductors will be modified. Thus, in this work we will construct the holographic superconductor models with $RF^{2}$ corrections. In order to extract the main physics and avoid the complex computation, we will use the probe approximation where the backreaction of matter fields on the metric can be neglected. By using the numerical shooting method \cite{HartnollPRL101} and the analytical Sturm-Liouville(S-L) method \cite{Siopsis}, we will obtain the relation between the $RF^2$ coupling parameter and the critical temperature as well as the condensation gap. Furthermore, we calculate the conductivity and discuss the effect of $RF^{2}$ corrections on the ratio $\omega_g/T_c$ in the gap frequency.

The organization of this work is as follows. In Sec. II, we will present the holographic superconductor model with $RF^2 $ corrections. In Sec. III, we will study the effect of $RF^{2}$ corrections on the condensation of the scalar operators. In Sec. IV, we will discuss the effect of $RF^{2}$ corrections on the conductivity. We will conclude in the last section of our main results.

\section{superconductor models with $RF^2 $ corrections}

We will introduce the model of holographic superconductor in the \emph{d}-dimensional Schwarzschild-AdS black hole background
\begin{eqnarray}\label{BH}
ds^2=-f(r)dt^{2}+\frac{dr^2}{f(r)}+r^{2}dx_{i}dx^{i},
\end{eqnarray}
with
\begin{eqnarray}
f(r)=r^2(1-\frac{r_{+}^{d-1}}{r^{d-1}}),
\end{eqnarray}
where $r_{+}$ is the black hole horizon and we have chosen units
such that the AdS radius is unity \cite{HorowitzPRD78}. We can
obtain the Hawking temperature
\begin{eqnarray}
T=\frac{(d-1)r_{+}}{4{\pi}},
\end{eqnarray}
which can be interpreted as the temperature of the CFT.

In order to construct the model of holographic superconductor with $RF^2$ corrections in the probe limit, we consider a Maxwell field and a charged complex scalar field coupled in the background of a
\emph{d}-dimensional spacetime via the action \cite{RCMyers,CaiPang}
\begin{eqnarray}\label{SWAction}
S=\int
d^{d}x\sqrt{-g}\left[-\frac{1}{4}F_{\mu\nu}F^{\mu\nu}+\mathcal{L}_{RF^{2}}-|\nabla\psi - iA\psi|^{2}
-m^2|\psi|^2 \right],
\end{eqnarray}
where $F_{\mu\nu}=\partial_{\mu}A_{\nu}-\partial_{\nu}A_{\mu}$ is
the field strength tensor and $m^{2}$ is the mass square of the
scalar field $\psi$. Taking the ansatz of the matter fields as
$\psi=\psi(r)$ and $A=\phi(r) dt$, we can obtain the equations of
motion for the scalar field $\psi$ and gauge field $\phi$
\begin{eqnarray}
\psi^{\prime\prime}+\left(
\frac{d-2}{r}+\frac{f^\prime}{f}\right)\psi^\prime
+\left(\frac{\phi^2}{f^2}-\frac{m^2}{f}\right)\psi=0\,,
\label{BHPsi}
\end{eqnarray}
\begin{eqnarray}
\left[1+4{\alpha}f\frac{(d-2)(d-3)}{r^{2}}\right]\phi^{\prime\prime}+\frac{d-2}{r}\left[1+4{\alpha}\frac{(d-3)}{r}\left(f^{\prime}+f\frac{d-4}{r}\right)\right]\phi^{\prime}
-\frac{2\psi^2}{f}\phi=0, \label{BHPhi}
\end{eqnarray}
where the prime denotes the derivative with respect to $r$.

At the horizon $r=r_{+}$, the regularity gives the boundary
conditions
\begin{eqnarray}
\psi(r_{+})=\frac{f^{\prime}(r_{+})\psi^{\prime}(r_{+})}{m^{2}},~~~~~~\phi(r_{+})=0.
\end{eqnarray}

Near the boundary $r\rightarrow\infty$, we get asymptotic behaviors
\begin{eqnarray}
\psi=\frac{\psi_{-}}{r^{\Delta_{-}}}+\frac{\psi_{+}}{r^{\Delta_{+}}}\,,\hspace{0.5cm}
\phi=\mu-\frac{\rho}{r^{d-3}}\,, \label{SWInfinity}
\end{eqnarray}
where $\Delta_\pm=\frac{1}{2}\left[(d-1)\pm\sqrt{(d-1)^{2}+4m^{2}}\right]$, $\mu$ and $\rho$ are interpreted the chemical potential and charge density in the dual field theory, respectively.
The coefficients $\psi_{-}$ and $\psi_{+}$ both multiply
normalizable modes of the scalar field equations and they correspond
to the vacuum expectation values
$\psi_{-}=\langle\mathcal{O}_{-}\rangle$,
$\psi_{+}=\langle\mathcal{O}_{+}\rangle$ of operators dual to the
scalar field according to the AdS/CFT correspondence. We can impose
boundary conditions that either $\psi_{-}$ or $\psi_{+}$ vanish
\cite{HartnollJHEP12,HartnollPRL101}. In the following discussion,
we will set \emph{d}=4 for simplicity and the coupling parameter
$\alpha$ is confined to the range $-1/20\leq \alpha
 \leq 1/4$ \cite{RCMyers}.

\section{The condensation of the scalar operators}

\subsection{The numerical method}

We will study the effects of the $RF^{2}$ coupling parameter $\alpha$ on the condensation of the
scalar operators by using the numerical shooting method. For concreteness, we will set $m^2L^2=-2$ in the following discuss. Actually, the other choices of the mass of the scalar field will not essentially change our results. We present the numerical results for the critical temperature $T_{c}$ by solving the equations of motion (\ref{BHPsi}) and (\ref{BHPhi}) in Table \ref{Tc}.

\begin{table}[ht]
\caption{\label{Tc} The critical temperature $T_{c}$ obtained by the numerical calculation for the
operators $\mathcal{O}_{-}$ and $\mathcal{O}_{+}$ with different values of $\alpha$ for $d=4$ and $m^2L^2=-2$. We have set $\rho=1$
in the table.}
\begin{tabular}{c c c c c c c c c}
         \hline
$\alpha$ &  -0.05 &  -0.01 &  0 &  0.01 &  0.05 &  0.1 &  0.2 & 0.25
        \\
        \hline
~~~~$T_c(\mathcal{O}_{-})$~~~~&~~~~$0.2055$~~~~&~~~~$0.2222$~~~~&
~~~~$0.2255$~~~~&~~~~$0.2286$~~~~&~~~~$0.2391$~~~~&~~~~$0.2494$~~~~&~~~~$0.2647$~~~~&~~~~$0.2707$~~~~
          \\
~~~~$T_c(\mathcal{O}_{+})$~~~~&~~~~$0.1049$~~~~&
~~~~$0.1162$~~~~&~~~~$0.1184$~~~~&
~~~~$0.1205$~~~~&~~~~$0.1278$~~~~&~~~~$0.1351$~~~~&~~~~$0.1460$~~~~&~~~~$0.1504$~~~~
          \\
        \hline
\end{tabular}
\end{table}

In Table \ref{Tc}, we can observe that the critical temperature $T_{c}$ for the operators $\mathcal{O}_{-}$ and $\mathcal{O}_{+}$ goes up as the coupling parameter $\alpha$ increases, which indicates that
the higher coupling parameter $\alpha$ make the condensation easier to be formed. This behavior is reminiscent of that seen for the holographic superconductor model with Weyl corrections where the critical temperature of a superconductor increases as we amplify the Weyl coupling
parameter \cite{Wu-Cao}. Thus, we conclude that the $RF^{2}$ corrections and the
Weyl corrections share some similar features for the condensation of the scalar
operators.

\begin{figure}[ht]
\includegraphics[scale=0.75]{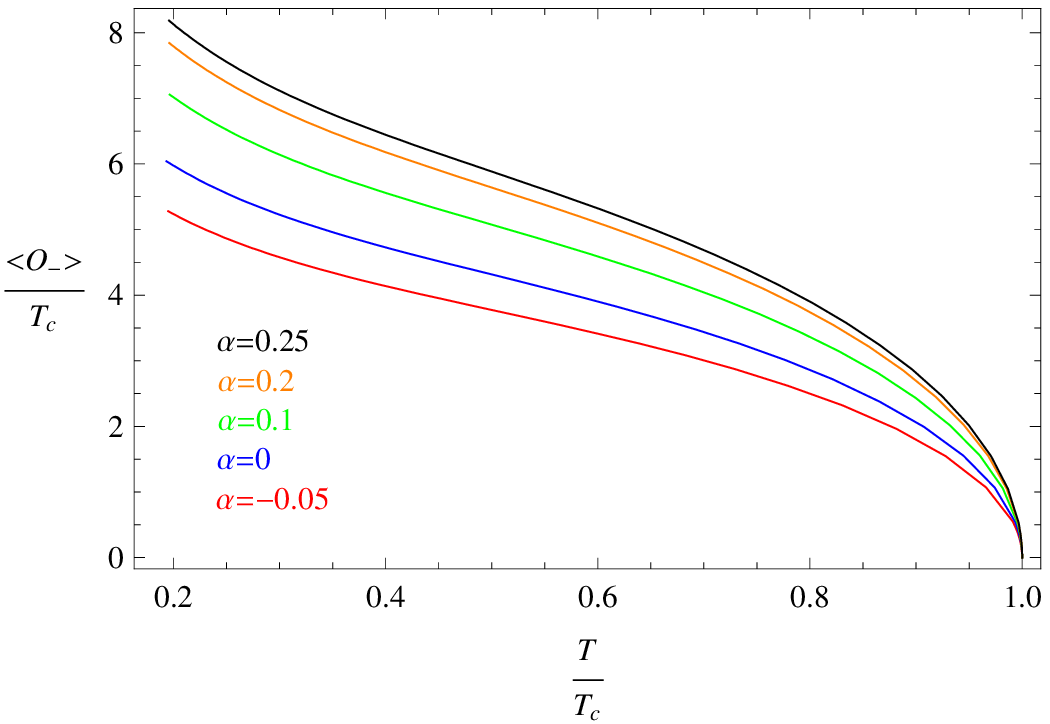}\vspace{0.0cm}
\includegraphics[scale=0.75]{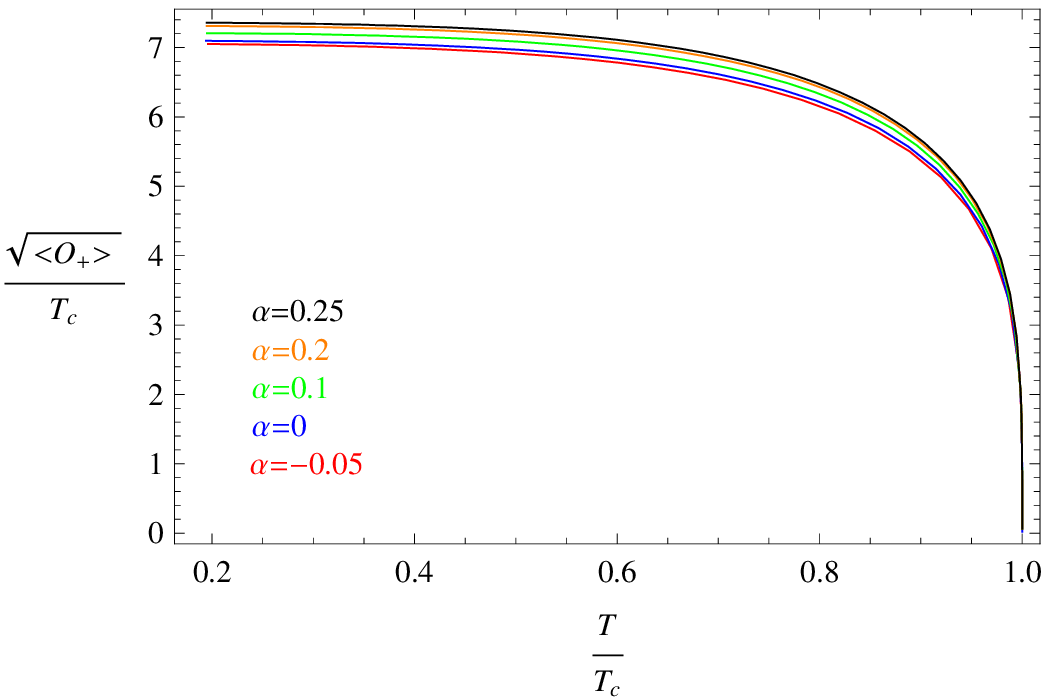}\\ \vspace{0.0cm}
\caption{\label{CondRF} (color online) The condensates of the
scalar operators $\mathcal{O}_{-}$ and $\mathcal{O}_{+}$ as a
function of temperature for the mass of the scalar field $m^2L^2=-2$
in $d=4$ dimension by using the numerical method. The five lines from bottom to top correspond to
increasing correction term, i.e., $\alpha=-0.05$ (red), $0$ (blue),
$0.1$ (green), $0.2$ (orange) and $0.25$ (black), respectively. }
\end{figure}

In order to study the behavior of the condensation further, in Fig. \ref{CondRF} we present the condensates of the scalar operators $\mathcal{O}_{-}$ and $\mathcal{O}_{+}$ as a function of temperature for the mass of the scalar field $m^2L^2=-2$ in $d=4$ dimension. From Fig. \ref{CondRF}, we can observe the curves in the right panel have similar behavior to the BCS theory for different $\alpha$, where the condensate goes to a constant at zero temperature. However, the curves for the operator $\mathcal{O}_{-}$ will diverge at low temperature, which are similar to that for the standard Maxwell electrodynamics in the probe limit neglecting backreaction of the spacetime \cite{HartnollPRL101}. Thus, we find that the holographic superconductors still exist even we consider $RF^{2}$ correction terms to the usual Maxwell electrodynamics.

From Fig. \ref{CondRF}, we see the higher correction term $\alpha$ makes the condensation gap larger for both scalar operators $\mathcal{O}_{-}$ and $\mathcal{O}_{+}$, which is completely different from the effects of the Weyl corrections discussed in \cite{Wu-Cao}. Consider that the $RF^{2}$ corrections and Weyl corrections are no longer equivalent when the background is at finite charge density, so our results may be natural and can be used to support the findings in Ref. \cite{CaiPang}.

\subsection{The analytical method}

In order to confirm the numerical results in the preceding subsection, we will study the condensation behavior by the analytical S-L method.

Introducing the variable $z=\frac{r_+}{r}$, we can rewrite the Eqs. (\ref{BHPsi}) and (\ref{BHPhi}) as
\begin{eqnarray}
\psi''+\frac{f^\prime}{f}\psi'+
\frac{r_+^{2}}{z^4}\left(\frac{\phi^{2}}{f^2}-\frac{m^2}{f}\right)\psi=0\label{NewBHPsi},
\end{eqnarray}
\begin{eqnarray}
\left(1+\frac{8\alpha z^2f}{r_+^2}\right)\phi''+\frac{8\alpha z}{r_{+}^2
}\left(2f+zf^\prime\right)\phi^\prime-\frac{2\psi^2r_+^{2}}{z^4f}\phi
=0\label{NewBHPhi},
\end{eqnarray}
where the prime now denotes derivative with respect to $z$. At the horizon $z=1$, the regularity gives the boundary conditions
\begin{eqnarray}
\psi'(1)=\frac{2}{3}\psi(1),~~~~~~\phi(1)=0.
\end{eqnarray}
Near the boundary ($z\rightarrow0$), the solutions behave like
\begin{eqnarray}
\psi(z)\approx
\langle\mathcal{O}_{i}\rangle\frac{z^{\Delta_i}}{r_+^{\Delta_i}}
,~~~~\phi(z) \approx \mu - \frac{\rho}{r_+}z\label{Solutionz0},
\end{eqnarray}
with $i=+$ or $i=-$.

At the critical temperature $T_c$, $\psi=0$, the Eq. (\ref{NewBHPhi}) reduces to
\begin{eqnarray}
\left(1+\frac{8\alpha z^2f}{r_{+c}^2}\right)\phi''+\frac{8\alpha z}{r_{+c}^2
}\left(2f+zf^\prime\right)\phi^\prime =0\label{NewBHPhipsi0}.
\end{eqnarray}
From the Eq. (\ref{NewBHPhipsi0}), we have
\begin{eqnarray}
\phi^\prime(z)=\frac{c_{1}}{1+8\alpha-8\alpha z^3}.
\end{eqnarray}
Using Eq. (\ref{Solutionz0}), we can get
\begin{eqnarray}
\phi^\prime(0)=-\frac{\rho}{r_{+c}}=\frac{c_{1}}{1+8\alpha}.
\end{eqnarray}
From above equation, we arrive at
\begin{eqnarray}
c_{1}=-\frac{(1+8\alpha)\rho}{r_{+c}}.
\end{eqnarray}
Therefore, we have
\begin{eqnarray}
\phi^\prime(z)=-\frac{(1+8\alpha)\lambda r_{+c}}{1+8\alpha-8\alpha
z^3}\approx -\lambda r_{+c}(1+\frac{8\alpha}{1+8\alpha}z^3),
\end{eqnarray}
where $\lambda=\rho/r_{+c}^2$ and $r_{+c}$ is the radius of the horizon at $T=T_c$. Integrating the above equation in the interval $[1,z]$, we have
\begin{eqnarray}
\phi(z)=\lambda r_{+c} \xi(\alpha,z)=\lambda
r_{+c}(1-z)\left[1+\frac{2\alpha}{1+8\alpha}(1+z)(1+z^2)\right].
\end{eqnarray}

As $T\rightarrow T_c$, the equation of $\psi$ becomes
\begin{eqnarray}
\psi''+\frac{h^\prime}{h}\psi'+ \frac{1}{z^4}\left(\frac{\lambda^2
\xi^2}{h^2}-\frac{m^2}{h}\right)\psi=0,
\end{eqnarray}
where we have defined a new function $h(z)=f(z)/r_+^2$. Near the boundary, we introduce a trial function $F(z)$ which
satisfies
\begin{eqnarray}
\psi(z)\sim\langle\mathcal{O}_{i}\rangle\frac{z^{\Delta_i}}{\sqrt 2
r_+^{\Delta_i}} F(z) \label{sol1},
\end{eqnarray}
with the boundary condition $F(0)=1$ and $F^\prime(0)=0$. Thus, we can get the equation of motion for $F(z)$
\begin{eqnarray}\label{SWFzmotion}
F^{\prime\prime}+\left(\frac{2\Delta_i}{z}+\frac{h'}{h}\right)
F^{\prime}+\left[\frac{\Delta_i(\Delta_{i}-1)}{z^2}+\frac{\Delta_i}{z}
\frac{h'}{h}+\frac{1}{z^{4}}\left(\frac{\lambda^2
\xi^2}{h^2}-\frac{m^{2}}{h}\right)\right]F=0.
\end{eqnarray}
Introducing a new function
\begin{eqnarray}
M(z)=z^{2\Delta_{i}-2}(z^{3}-1),
\end{eqnarray}
we can rewrite Eq. (\ref{SWFzmotion}) as
\begin{eqnarray}\label{NSWFzmotion}
(MF^{\prime})^{\prime}+\left[\frac{\Delta_i(\Delta_{i}-1)}{z^2}+\frac{\Delta_i}{z}
\frac{h'}{h}+\frac{1}{z^{4}}(\frac{\lambda^2
\xi^2}{h^2}-\frac{m^{2}}{h})\right]MF=0.
\end{eqnarray}
Following the Sturm-Liouville eigenvlaue problem
\cite{Gelfand-Fomin}, we deduce the expression which can be used to
estimate the minimum eigenvalue of $\lambda^2$
\begin{eqnarray}\label{SWeigenvalue}
\lambda^{2}=\frac{\int^{1}_{0}M\left(F'^{2}-PF^{2}\right)dz}{\int^{1}_{0}WF^{2}dz},
\end{eqnarray}
with
\begin{eqnarray}
&&P=\frac{\Delta_{i}(\Delta_{i}-1)}{z^{2}}+\frac{\Delta_{i}}{z}\frac{h'}{h}-\frac{m^{2}}{z^{4}h},\nonumber\\
&&W=\frac{\xi^2}{z^{4}h^2}M.
\end{eqnarray}
We assume the trial function to be $F(z)=1-az^{2}$, where $a$ is a
constant.

With Eq. (\ref{SWeigenvalue}), we can get the value of $T_c$ for different $\alpha$. For example, we consider the $\mathcal{O_-}$ and $m^2L^2=-2$, $T_c=0.225\sqrt{\rho}$ and $a=0.2389$ for $\alpha=0$,
$T_c=0.2279\sqrt{\rho}$ and $a=0.2375$ for $\alpha=0.01$, $T_c=0.2362\sqrt{\rho}$ and $a=0.2332$ for $\alpha=0.05$. In Table \ref{SLTc} we list the results of the critical temperature $T_{c}$ for different $\alpha$ by using the S-L method. Comparing with numerical results given in Table \ref{Tc}, we find that the analytic results derived from S-L method agree well with the numerical calculation.

\begin{table}[ht]
\caption{\label{SLTc} The critical temperature $T_{c}$ obtained by the analytical S-L method for the
operators $\mathcal{O}_{-}$ and $\mathcal{O}_{+}$ with different values of $\alpha$ for $d=4$ and $m^2L^2=-2$. We have set $\rho=1$ in the table.}

\begin{tabular}{c c c c c c c c c}
         \hline
$\alpha$ &  -0.05 &  -0.02 &  -0.01 &  0  &  0.01 &  0.02 &  0.05 &
0.1
        \\
        \hline
~~~~$T_c(\mathcal{O}_{-})$~~~~&~~~~$0.1969$~~~~&~~~~$0.2172$~~~~&~~~~$0.2214$~~~~&
~~~~$0.2250$~~~~&~~~~$0.2279$~~~~&~~~~$0.2304$~~~~&~~~~$0.2362$~~~~&~~~~$0.2423$~~~~
          \\
~~~~$T_c(\mathcal{O}_{+})$~~~~&~~~~$0.1000$~~~~&~~~~$0.1115$~~~~&
~~~~$0.1146$~~~~&~~~~$0.1170$~~~~&
~~~~$0.1191$~~~~&~~~~$0.1209$~~~~&~~~~$0.1249$~~~~&~~~~$0.1291$~~~~
          \\
        \hline
\end{tabular}
\end{table}

Now we are in a position to investigate the critical phenomena and obtain the critical exponent of the system. Away from (but close to) the critical temperature, using the
Eq. (\ref{sol1}) and $h(z)=f(z)/r_+^2$, we can rewrite the Eq. (\ref{NewBHPhi}) as
\begin{eqnarray}
(1+8\alpha z^2h)\phi''+8\alpha z
(2h+zh^\prime)\phi^\prime=\frac{\langle\mathcal{O}_{i}\rangle^2}{r_+^{2{\Delta_i}}}\frac{z^{2{\Delta_i}-4}F^2}{h
}\phi.
\end{eqnarray}
Consider that the condensation of the scalar operator $\mathcal{O}_{i}$ is so small when $T\rightarrow T_{c}$, we can expand $\phi(z)$ as
\begin{eqnarray}
\frac{\phi(z)}{r_+}=\lambda
\xi+\frac{\langle\mathcal{O}_{i}\rangle^2}{r_+^{2{\Delta_i}}}\chi(z).
\end{eqnarray}
Thus, we can get the equation of motion for $\chi$ near the critical
temperature
\begin{eqnarray}
\chi''-24\alpha
z^2\chi'=\lambda\frac{z^{2{\Delta_i}-4}F^2\xi}{h(1+8\alpha z^2h)},
\end{eqnarray}
with $\chi(1)=0$ and $\chi'(1)=0$. Defining a function
\begin{eqnarray}
V=e^{-8\alpha z^3},
\end{eqnarray}
we have
\begin{eqnarray}
(V\chi')'=\lambda e^{-8\alpha
z^3}\frac{z^{2{\Delta_i}-4}F^2\xi}{h(1+8\alpha z^2h)}.
\end{eqnarray}
Integrating above equation from $z=0$ to $z=1$, we have
\begin{eqnarray}
\chi'(0)=-\lambda Q=-\lambda\int_0^1 e^{-8\alpha
z^3}\frac{z^{2{\Delta_i}-4}F^2\xi}{h(1+8\alpha z^2h)}dz.
\end{eqnarray}

We expand $\phi(z)$ near $z=0$
\begin{eqnarray}
\frac{\phi}{r_+}=\frac{\mu}{r_+}-\frac{\rho z}{r_+^2}=\lambda
\xi+\frac{\langle\mathcal{O}_{i}\rangle^2}{r_+^{2{\Delta_i}}}\chi(z)=\lambda
\xi+\frac{\langle\mathcal{O}_{i}\rangle^2}{r_+^{2{\Delta_i}}}\left[\chi(0)+z\chi'(0)+...\right].
\end{eqnarray}
From the coefficients of the $z$ term, we can get
\begin{eqnarray}
\frac{\rho}{r_+^2}=\lambda-\frac{\langle\mathcal{O}_{i}\rangle^2}{r_+^{2{\Delta_i}}}\chi'(0)
=\lambda\left(1+\frac{\langle\mathcal{O}_{i}\rangle^2}{r_+^{2{\Delta_i}}}Q\right).
\end{eqnarray}
When $T\rightarrow T_c$, we can express $\langle\mathcal{O}_{i}\rangle$ as
\begin{eqnarray}
\frac{\langle\mathcal{O}_{i}\rangle^{\frac{1}{\Delta_i}}}{T_c}
=\left(\frac{4\sqrt{2}\pi}{3\sqrt{Q}}\sqrt{1-\frac{T}{T_c}}\right)^{\frac{1}{\Delta_i}},
\end{eqnarray}
which shows that the phase transition belongs to the second order and the critical exponent of the system takes the mean-field value $1/2$. The $RF^{2}$ corrections will not influence the result, which agrees well with the numerical finding shown in Fig. \ref{CondRF}.

\begin{figure}[ht]
\includegraphics[scale=0.75]{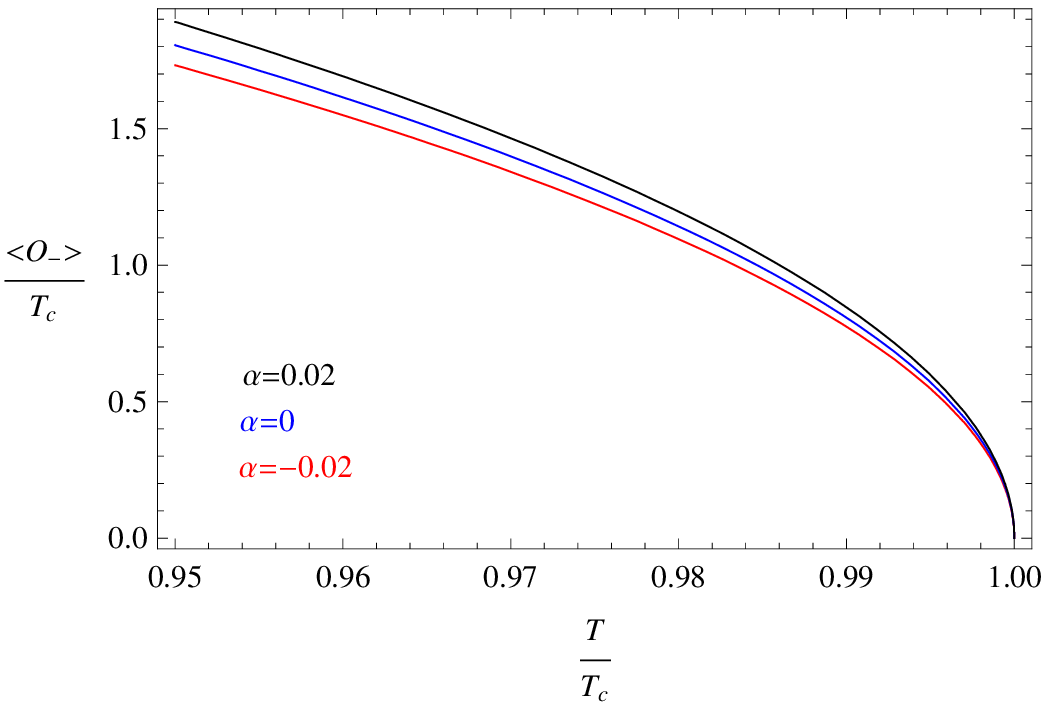}\vspace{0.0cm}
\includegraphics[scale=0.75]{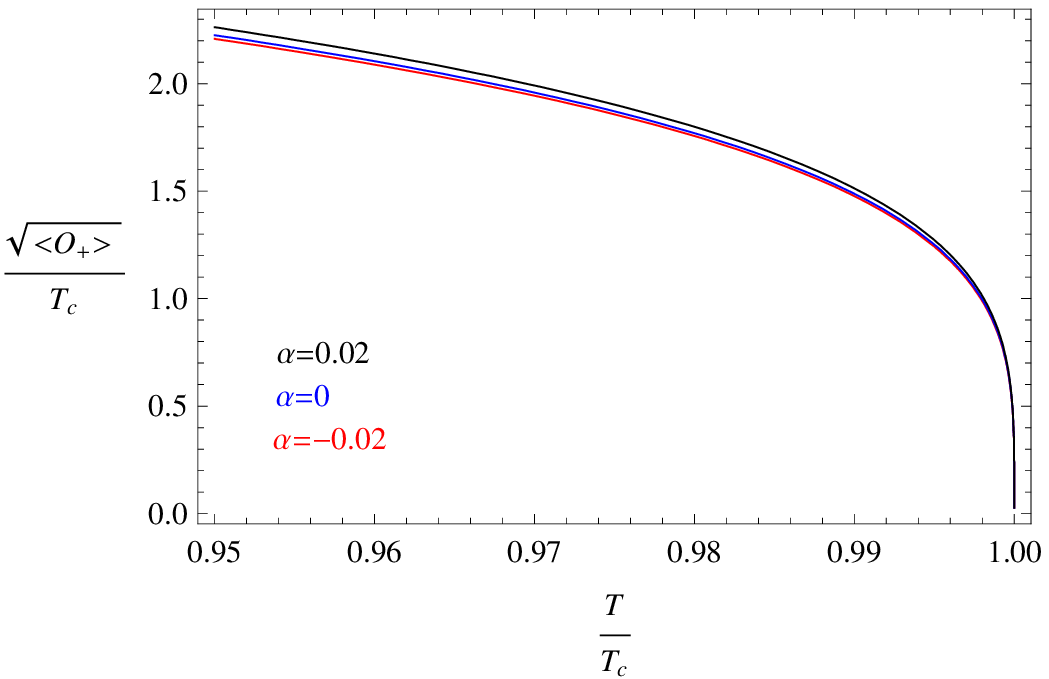}\\ \vspace{0.0cm}
\caption{\label{SLCondCMF} (color online) The condensates of the
scalar operators $\mathcal{O}_{-}$ and $\mathcal{O}_{+}$ as a
function of temperature for the mass of the scalar field $m^2L^2=-2$
in $d=4$ dimension by using the analytical method. The three lines from bottom to top correspond to
increasing correction term, i.e., $\alpha=-0.02$ (red), $0$ (blue),
$0.02$ (black), respectively. }
\end{figure}

Using the analytical method, we plot the condensates of the scalar operators $\mathcal{O}_{-}$ and $\mathcal{O}_{+}$ as a function of temperature for the mass of the scalar field $m^2L^2=-2$ with $d=4$ dimension in Fig. \ref{SLCondCMF}. From this figure, we can find that the gap of the
condensation goes up with the increase of $\alpha$ for both scalar operators $\mathcal{O}_{-}$ and $\mathcal{O}_{+}$, which is also good agreement with the numerical results obtained in Fig. \ref{CondRF}.

\section{Conductivity}

In this section, we will calculate the effect of the $RF^2$ coupling parameter $\alpha$ on the conductivity $\sigma$. In order to study the conductivity, we introduce the perturbed
Maxwell field $\delta A_{x}=A_{x}(r)e^{-i\omega t}dx$ and get the equation of motion for $\delta A_{x}$
\begin{eqnarray}
&A_{x}^{\prime\prime}&+\left\{\frac{d-4}{r}+\frac{f^\prime}{f}+
4\alpha(d-3)\left[-\frac{2(d-4)}{r^{3}}f+\frac{d-5}{r^{2}}f^{\prime}+\frac{f^{\prime\prime}}{r}\right]
\left[1+4\alpha(d-3)\left(\frac{d-4}{r^{2}}f+\frac{f^{\prime}}{r}\right)\right]^{-1}\right\}A_{x}^\prime
\nonumber\\
&&+\left\{\frac{\omega^2}{f^2}-\frac{2\psi^{2}}{f}
\left[1+4\alpha(d-3)\left(\frac{d-4}{r^{2}}f+\frac{f^{\prime}}{r}\right)\right]^{-1}\right\}A_{x}=0,
\label{Conductivity Equation}
\end{eqnarray}
which can be used to calculate the conductivity. We also focus on the $d=4$ for simplicity. The ingoing wave boundary condition near the horizon is the same to the case of the standard holographic
superconductor models \cite{HartnollPRL101,HartnollJHEP12}
\begin{eqnarray}
A_{x}(r)\sim f(r)^{-\frac{i\omega}{3r_+}},
\end{eqnarray}
and at the asymptotic AdS boundary $r\rightarrow\infty$, the solutions reduce to
\begin{eqnarray}
A_{x}=A^{(0)}+\frac{A^{(1)}}{r}.
\end{eqnarray}
We can get the conductivity of the dual superconductor

\begin{eqnarray}\label{CMFConductivity}
\sigma=-\frac{iA^{(1)}}{\omega A^{(0)}}\ .
\end{eqnarray}
In the following discussion, we also fix the scalar mass $m^2L^2=-2$ and calculate the conductivity for
different coupling parameter $\alpha$ by using the shooting method.

\begin{figure}[ht]
\includegraphics[scale=0.51]{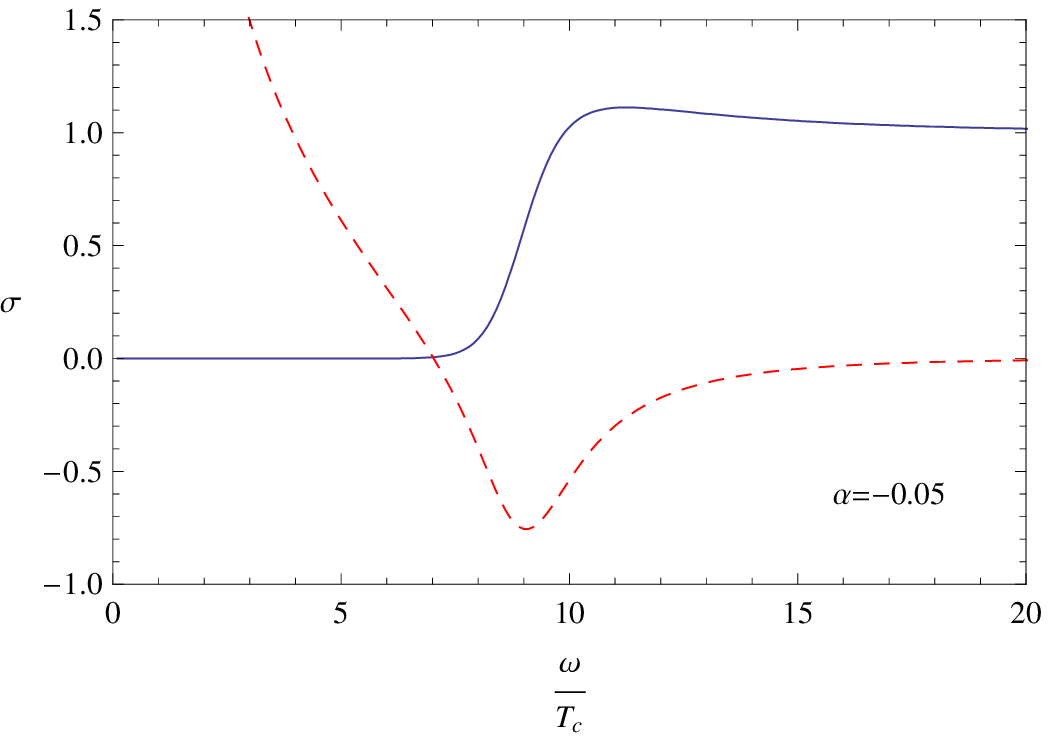}\hspace{0.2cm}%
\includegraphics[scale=0.51]{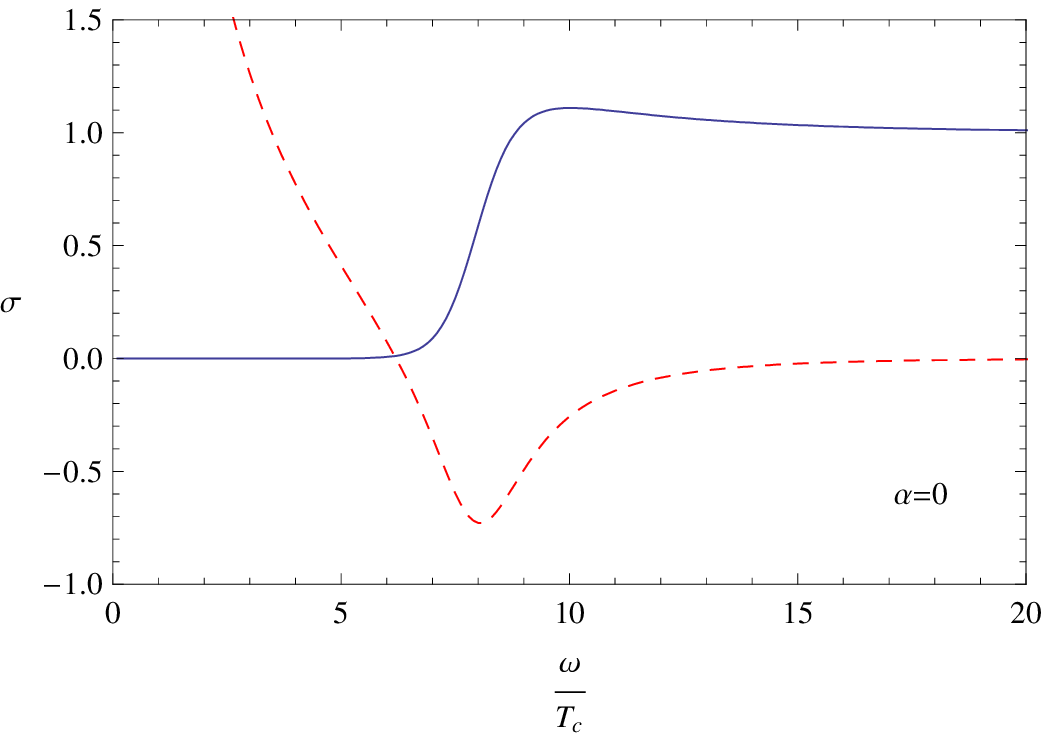}\vspace{0.0cm}
\includegraphics[scale=0.51]{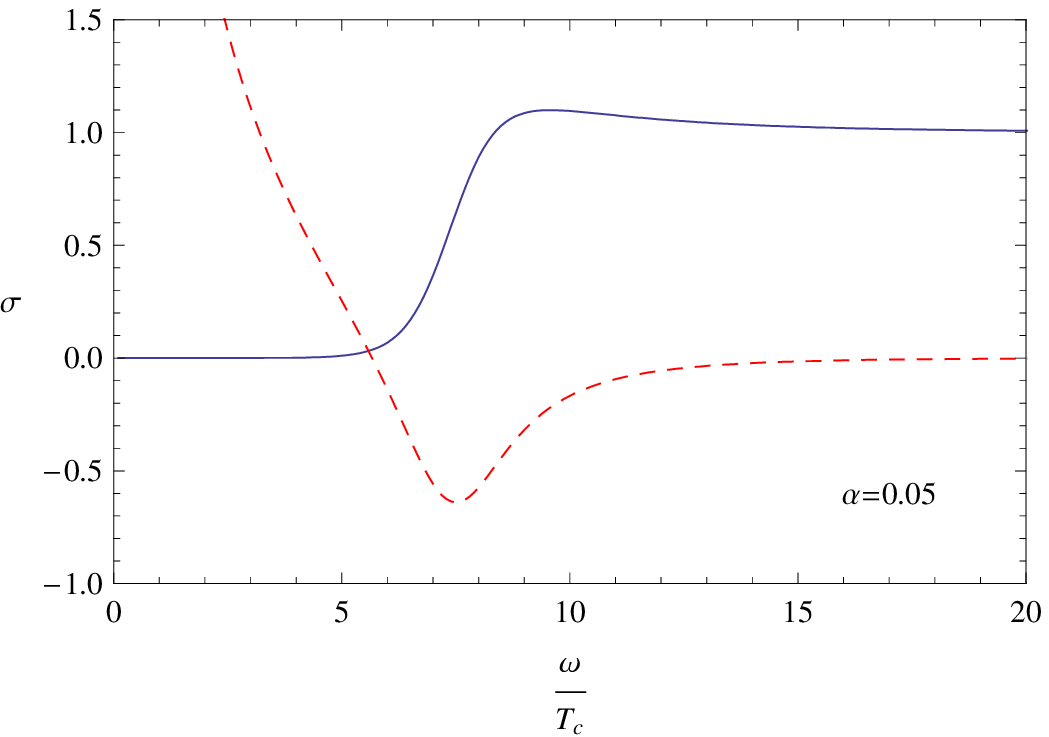}\\ \hspace{0.2cm}%
\includegraphics[scale=0.51]{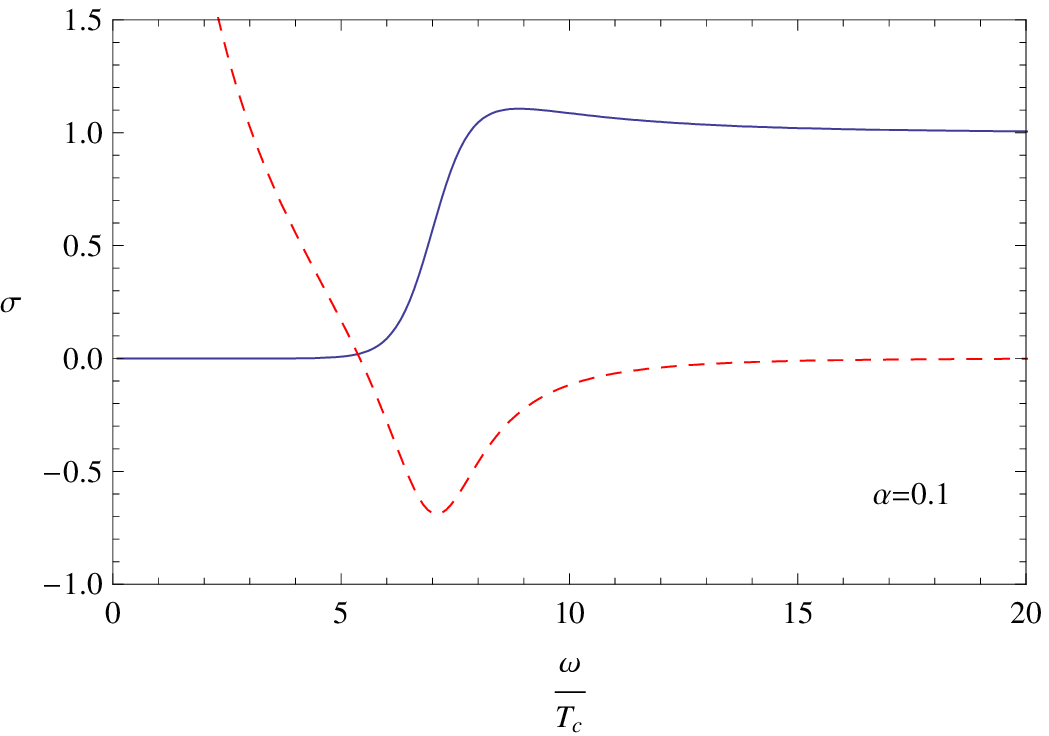}\vspace{0.0cm}
\includegraphics[scale=0.51]{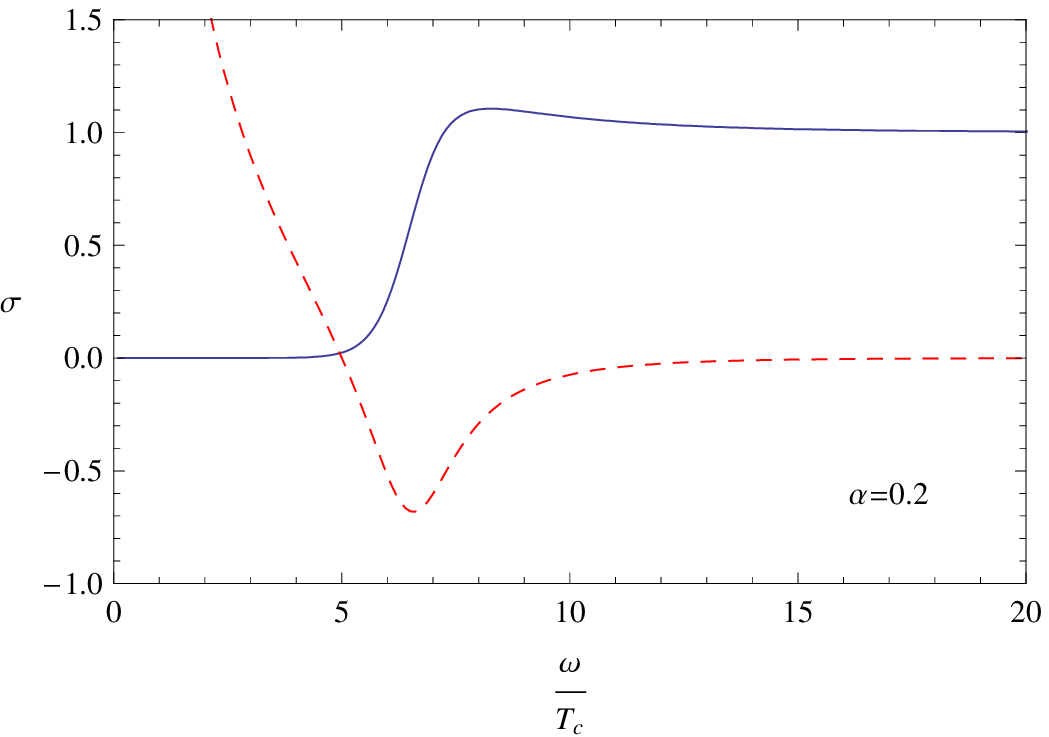}\hspace{0.2cm}%
\includegraphics[scale=0.51]{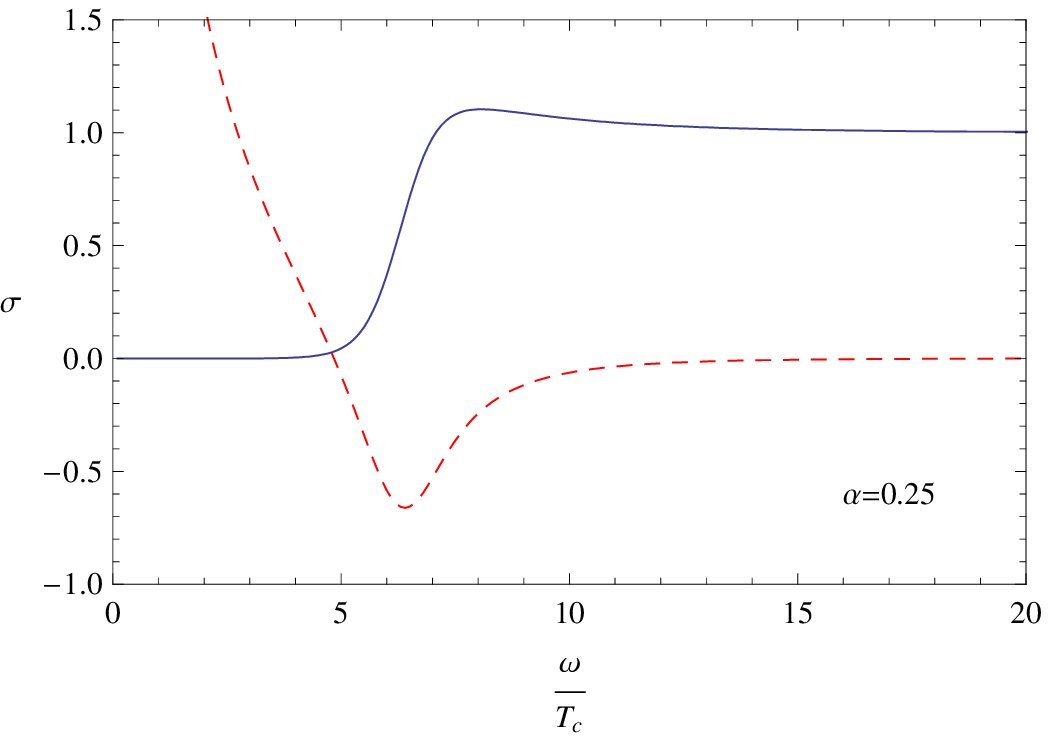}\\ \vspace{0.0cm}
\caption{\label{RFConductivity} (color online) Conductivity of
($2+1$)-dimensional superconductors with the $RF^{2}$ corrections
for the fixed mass of the scalar field $m^2L^2=-2$ and different correction
terms, i.e., $\alpha=-0.05, 0, 0.05, 0.1, 0.2$ and $0.25$.}
\end{figure}

In Fig. \ref{RFConductivity} we plot the frequency dependent
conductivity obtained by solving the Maxwell equation
(\ref{Conductivity Equation}) numerically for $\alpha=-0.05$, $0$,
$0.05$, $0.1$, $0.2$ and $0.25$ at temperatures $T/T_{c}\approx0.2$.
The blue (solid) line and red (dashed) line represent the real part
and imaginary part of the conductivity $\sigma(\omega)$
respectively. For the same mass of the scalar field, we find that
with the increase of the $RF^{2}$ coupling parameter $\alpha$, the
gap frequency $\omega_{g}$ becomes smaller. As the $RF^{2}$ coupling
parameter increases the value of $\omega_g/T_c$ decreases, which is
similar to the Weyl corrections. Meanwhile, for increasing $RF^{2}$ correction term, we have larger deviations from the value $\omega_g/T_c\approx 8$, which shows that the high $RF^{2}$ corrections really change the expected universal relation in the gap frequency \cite{HorowitzPRD78}, which is similar to the effect of the Weyl corrections \cite{Wu-Cao}.

\section{Conclusions}

We have introduced the holographic superconductors with $RF^2$ corrections in the probe limit in order to understand the influences of the $1/N$ or $1/\lambda$ corrections on the holographic dual model in the AdS black hole background. Similar to the Weyl correction, we found that with the increase of the $RF^2$ correction the critical temperature becomes larger, which shows that the higher correction term can make the condensation easier to form. However, we observed that the condensation gap becomes larger as the $RF^{2}$ correction term increases, which is in contrast to the effect of Weyl correction. We confirmed our numerical result by using the Sturm-Liouville analytic method and concluded that the $RF^2$ correction and the Weyl correction do play different roles if a non-trivial hair for the black hole has been triggered. Our results may be natural and can be used to support the findings in Ref. \cite{CaiPang} since the $RF^{2}$ corrections and Weyl corrections are no longer equivalent when the background is at finite charge density. We also investigated the conductivity and found that the higher $RF^2$ correction term will result in the larger deviations from the universal value $\omega_g/T_c\approx 8$ for the gap frequency, which is similar to the effect of the Weyl correction \cite{Wu-Cao}.

{\bf Note added}------While we were completing this paper, a complementary paper of \"{O}. Sert and M. Adak \cite{SertAdak} on non-minimal $RF^2$-type corrections to holographic superconductor appeared in arXiv.

\begin{acknowledgments}

This work was supported by the National Natural Science Foundation of China under Grant Nos. 11275066 and 11175065; the National Basic Research of China under Grant No. 2010CB833004; PCSIRT under Grant No. IRT0964; NCET under Grant No.10-0165; Hunan Provincial Natural Science Foundation of China under Grant Nos. 12JJ4007 and 11JJ7001; and the Construct Program of the National Key Discipline.

\end{acknowledgments}


\begin{thebibliography}{99}

\bibitem{Maldacena}
J. Maldacena, Adv. Theor. Math. Phys. {\bf 2}, 231 (1998) [Int. J.
Theor. Phys. {\bf 38}, 1113 (1999)].

\bibitem{Gubser1998}
S.S. Gubser, I.R. Klebanov, and A.M. Polyakov, Phys. Lett. B {\bf
428}, 105 (1998).

\bibitem{Witten}
E. Witten, Adv. Theor. Math. Phys. {\bf 2}, 253 (1998).

\bibitem{GubserPRD78}
S.S. Gubser, Phys. Rev. D {\bf 78}, 065034 (2008).

\bibitem{HartnollJHEP12}
S.A. Hartnoll, C.P. Herzog, and G.T. Horowitz, J. High Energy Phys. {\bf 12}, 015 (2008).

\bibitem{HartnollPRL101}
S.A. Hartnoll, C.P. Herzog, and G.T. Horowitz, Phys. Rev. Lett. {\bf
101}, 031601 (2008).

\bibitem{HartnollRev}
S.A. Hartnoll, Class. Quant. Grav. {\bf 26}, 224002 (2009).

\bibitem{HerzogRev}
C.P. Herzog, J. Phys. A {\bf 42}, 343001 (2009).

\bibitem{HorowitzRev}
G.T. Horowitz, arXiv: 1002.1722 [hep-th].

\bibitem{Gregory}
R. Gregory, S. Kanno, and J. Soda, J. High Energy Phys. {\bf 10},
010 (2009).

\bibitem{Pan-Wang}
Q.Y. Pan, B. Wang, E. Papantonopoulos, J. Oliveria, and A.B. Pavan,
Phys. Rev. D {\bf 81}, 106007 (2010).

\bibitem{Ge-Wang}
X.H. Ge, B. Wang, S.F. Wu, and G.H. Yang, J. High Energy Phys. {\bf
08}, 108 (2010); arXiv:1002.4901 [hep-th].

\bibitem{JS2010}
J.L. Jing and S.B. Chen, Phys. Lett. B {\bf 686}, 68 (2010).

\bibitem{JLQS2012}
J.L. Jing, L.C. Wang, Q.Y. Pan, and S.B. Chen, Phys. Rev. D {\bf 83}, 066010 (2012).

\bibitem{PJWPRD}
Q.Y. Pan, J.L. Jing, and B. Wang, Phys. Rev. D {\bf 84}, 126020 (2011).

\bibitem{JingJHEP}
J.L. Jing, Q.Y. Pan, and S.B. Chen, J. High Energy Phys. {\bf 11}, 045 (2011).

\bibitem{SDSL2012}
S. Gangopadhyay and D. Roychowdhury, J. High Energy Phys. {\bf 05}, 002 (2012);  {\bf 05}, 156 (2012); {\bf 08}, 104 (2012).

\bibitem{LPW2012}
Y.Q. Liu, Y. Peng, and B. Wang, arXiv:1202.3586 [hep-th].

\bibitem{BGRL2012}
R. Banerjee, S. Gangopadhyay, D. Roychowdhury, and A. Lala, arXiv:1208.5902 [hep-th].

\bibitem{JPCPLB}
J.L. Jing, Q.Y. Pan, and S.B. Chen, Phys. Lett. B {\bf 716}, 385 (2012).

\bibitem{Roychowdhury}
D. Roychowdhury, Phys. Lett. B {\bf 718}, 1089 (2013).

\bibitem{ZPCJ2012}
Z.X. Zhao, Q.Y. Pan, S.B. Chen, and J.L. Jing, arXiv:1212.6693 [hep-th].

\bibitem{Wu-Cao}
J.P. Wu, Y. Cao, X.M. Kuang, and W.J. Li, Phys. Lett. B {\bf 697}, 153 (2011).

\bibitem{MaCW}
D.Z. Ma, Y. Cao, and J.P. Wu, Phys. Lett. B {\bf 704}, 604 (2011).

\bibitem{DMMPLA}
D. Momeni and M.R. Setare, Mod. Phys. Lett. A {\bf 26}, 2889 (2011).

\bibitem{RoychowdhuryPRD}
D. Roychowdhury, Phys. Rev. D {\bf 86}, 106009 (2012).

\bibitem{MomeniWeyl}
D. Momeni, N. Majd, and R. Myrzakulov, Europhys. Lett. {\bf 97}, 61001 (2012); Int. J. Mod. Phys. A {\bf 27}, 1250128 (2012).

\bibitem{ZPJ2012}
Z.X. Zhao, Q.Y. Pan, and J.L. Jing, arXiv:1212.3062 [hep-th].

\bibitem{RCMyers}
R.C. Myers, S. Sachdev, and A. Singh, Phys. Rev. D {\bf 83}, 066017 (2011).

\bibitem{CaiPang}
R.G. Cai and D.W. Pang, Phys. Rev. D {\bf 84}, 066004 (2011).

\bibitem{Siopsis}
G. Siopsis and J. Therrien, J. High Energy Phys. {\bf 05}, 013
(2010).

\bibitem{HorowitzPRD78}
G.T. Horowitz and M.M. Roberts, Phys. Rev. D {\bf 78}, 126008
(2008).

\bibitem{Gelfand-Fomin}
I.M. Gelfand and S.V. Fomin, \textit{Calculaus of Variations},
Revised English Edition, Translated and Edited by R.A. Silverman,
Prentice-Hall, Inc. Englewood Cliff, New Jersey (1963).

\bibitem{SertAdak}
\"{O}. Sert and M. Adak, arXiv:1301.3328 [hep-th].

\end{thebibliography}
\end{document}